\documentstyle[12pt,epsf]{article}

\begin{document}
\hfill {\small \bf TSU--HEPI--99--01}\\
\vglue -0.5cm
\hfill {\small \bf UA/NPPS--6--99}\\
\vglue -0.5cm
\hfill {\small \bf hep-ph/9911506}

\begin{center} \begin{bf}
POLARIZED HEAVY QUARK PHOTOPRODUCTION \\
        IN NEXT TO LEADING ORDER 
\end{bf} \end{center}

{\begin{center}
Z.V.~Merebashvili$^{*}$  
\end{center}}
\vglue -0.35cm
\parbox{6.4in}{\leftskip=1.0pc
{Department of Physics, McGill University, Montreal,
H3A 2T8, Canada} }

\vglue 0.25cm
{\begin{center}
A.P.~Contogouris$^{**}$,
              G.~Grispos  
\end{center}}
\vglue -0.35cm
\parbox{6.4in}{\leftskip=1.0pc
{Nuclear and Particle Physics, University of Athens,
Athens 15771, Greece} }

\renewcommand{\thefootnote}{*}
\footnotetext{Present address: Tbilisi
             State University, Tbilisi 380086, Georgia}     
\renewcommand{\thefootnote}{**}
\footnotetext{also at Department of Physics,
                        McGill University, Montreal, H3A2T8, Canada}
\vglue 0.25cm
{\small
Results of our next-to-leading order analytic calculation for
longitudinally polarized photoproduction of heavy quarks
are presented. Certain differential cross sections and the corresponding
asymmetries are computed 
for energy ranges accessible to CERN and HERA. We argue that determination of
the polarized gluon distribution is possible at both
places: at HERA certain acceptance cuts are necessary.
}

\vglue 0.5cm
Deep Inelastic Scattering (DIS) of longitudinally polarized particles has
been very useful in extracting the polarized valence distributions.
However, it is quite limited in extracting the polarized gluon
distribution $\Delta g$. Much was already said at this Workshop about the 
importance
of knowing $\Delta g$, and there is no need to talk about it any further.
In order to extract $\Delta g$ with a good accuracy one needs experiments
involving longitudinally polarized reactions dominated by subprocesses
with initial gluons. 

One such reaction is the inclusive polarized photoproduction of heavy
quarks:
\begin{equation}
\label{process}
\vec{\gamma}+\vec{p}\rightarrow Q \bar{(Q)} + X,
\end{equation}
which at leading order (LO) proceeds through the subprocess:
\begin{equation}
\label{lo}
\vec{\gamma}+\vec{g}\rightarrow Q + \bar{Q}
\end{equation}
There are several experimental proposals on the subject in
various stages of approval \cite{prop}.

Next-to-leading order corrections (NLOC) are very important for several
reasons: a) The leading order estimates depend strongly on the a priory
unknown renormalization and factorization scales; in general the NLOC 
reduce this uncertainty. More precisely, going to one level higher in
the strong coupling $\alpha_s$ may reduce the uncertainty by order
$\alpha_s$; b) NLOC may be large and negative in some regions
of phase space, making cross sections small and difficult to measure;
c) At NLO, subprocesses unrelated to the LO ones could come into play. In principle,
the contribution of such subprocesses could be significant in certain
important parts of phase space and this could affect even the shape
of the final cross sections. However, admittedly in the present case, the
subprocesses
\begin{equation}
\label{gamq}
\vec{\gamma}+\vec{q}(\vec{\bar{q}})\rightarrow Q + \bar{Q}
+ q (\bar{q}), 
\end{equation}
($q=$light quark) which belong to this class, make a relatively small contribution.
The same holds for the corresponding unpolarized subprocesses \cite{SN},
as well as for other such subprocesses (taken separately) we have studied \cite{CKM}.

Nevertheless, irrespective of relative sizes, this talk presents a \textit{complete}
NLO calculation of polarized photoproduction of heavy quarks. Some results, in a very
preliminary stage, have already
been reported \cite{old}.

An independent calculation of NLOC has appeared \cite{BS}, 
and we subsequently comment on it.

Let $M_{i}(\lambda_1,\lambda_2)$ be the amplitude of any of the
contributing subprocesses and $\lambda_1, \lambda_2$ the helicities
of the initial partons. Then unpolarized and polarized, cross sections 
correspond to the quantities:
\begin{equation}
\label{sigma}
\frac{1}{2} \Sigma [M_{i}(++) \stackrel{\ast}{M}_{j}(++) \pm
M_{i}(+-) \stackrel{\ast}{M}_{j}(+-)],
\end{equation}
where $\Sigma$ denotes summation over the helicities and colors
of the final particles and average over the colors of the
initial. Note for the unpolarized case one usually  averages over $n-2$ spin
degrees of freedom for every incoming boson, which amounts to multiplying
(\ref{sigma}) by $2/(n-2)$ for each initial boson.

Note that the Abelian part of the NLOC to our dominant subprocess (\ref{lo})
is of interest in itself, for it determines the NLOC to
$\vec{\gamma}+\vec{\gamma} \rightarrow Q(\bar{Q})+X$ \cite{{cmc},{JT}}.
Note also that, as it was already done in \cite{cmc}, a calculation of  NLOC
for a polarized reaction and the corresponding unpolarized one can
be done by simply using different projection operators.

Our regularization scheme is dimensional reduction (RD) properly
supplemented by finite counterterms \cite{cmc} and conversion terms
\cite{MCG} (see also below). We work in the Feynman gauge.

In determining the loop contributions, the renormalization
of the heavy quark mass and wave function were carried on
shell. Furthermore, in our scheme, the contribution of a heavy quark loop
in the gluon self-energy is subtracted out, i.e. the 
heavy quark is decoupled. This is consistent with parton
distributions of which the evolution
is determined from split functions involving only light quarks.

Our loop contributions have been determined by REDUCE \cite{reduce}
and to some extent by FORM \cite{form}. 
Tensor integrals have been reduced to 
scalar ones. It is worth noting that, in general, the coefficients of pole terms 
$1/\varepsilon^2$ and $1/\varepsilon$ for separate graphs are not 
proportional to the LO term; only their sums are.

Gluon Bremsstrahlung contributions of e.g.
$\vec{\gamma}+\vec{p}\rightarrow Q+X$ are determined, as usual,
by working in the c.m. frame of $\bar{Q}$ and $g$ (Gottfried-Jackson).
Many necessary $n$- and 4-dimensional integrals are given in
\cite{been} and the rest were derived by ourselves \cite{MCG}. 
Factorization terms are added in the usual way, and the cancellation of the loop  
$1/\varepsilon^2$ and $1/\varepsilon$ singularities is a
good indication of the correctness of our results.

Here we leave out a resolved $\gamma$ component; this, as well as 
an estimate of the anticipated experimental errors, is given in \cite{MCG}.

Our cross sections are convoluted with parton distributions evolved
via two-loop split functions. However, polarized split functions have
been determined in the modified t'Hooft-Veltman (HV) scheme \cite{hvbm}.
Thus, conversion terms were necessary to transform our
results to the HV scheme. 

We now present some numerical results for the polarized differential
cross sections: $\Delta d\sigma/dp_T$ and $\Delta d\sigma/dy$, where $p_T$ and $y$
the transverse momentum and c.m. rapidity with respect to the photon
of a heavy quark of mass $m$. 
For the renormalization ($\mu$) and factorization (M)
scales we take: $\mu^2=M^2=p_T^2+m^2$ for $\Delta d\sigma/dp_T$ and 
$\mu^2=M^2=4\,m^2$ for $\Delta d\sigma/dy$. With $\sqrt S$ the c.m.
energy of the colliding particles we define
$x_T=2\,p_T/\sqrt{S}$. We use the three sets of polarized parton distributions
(pd) of \cite{gs}, which mainly differ in $\Delta g$: Sets A and B differ in the size
of $\Delta g$, set C also in the shape. Our asymmetries are defined as follows:
\begin{equation}
\label{asymm}
A_{LL} \left( p_T \right)=\frac{\Delta d\sigma/dp_T}{d\sigma/dp_T}, \qquad
A_{LL} \left( y \right)=\frac{\Delta d\sigma/dy}{d\sigma/dy} 
\end{equation}
\begin{figure}
\centering
\mbox{\epsfysize=86mm\epsffile{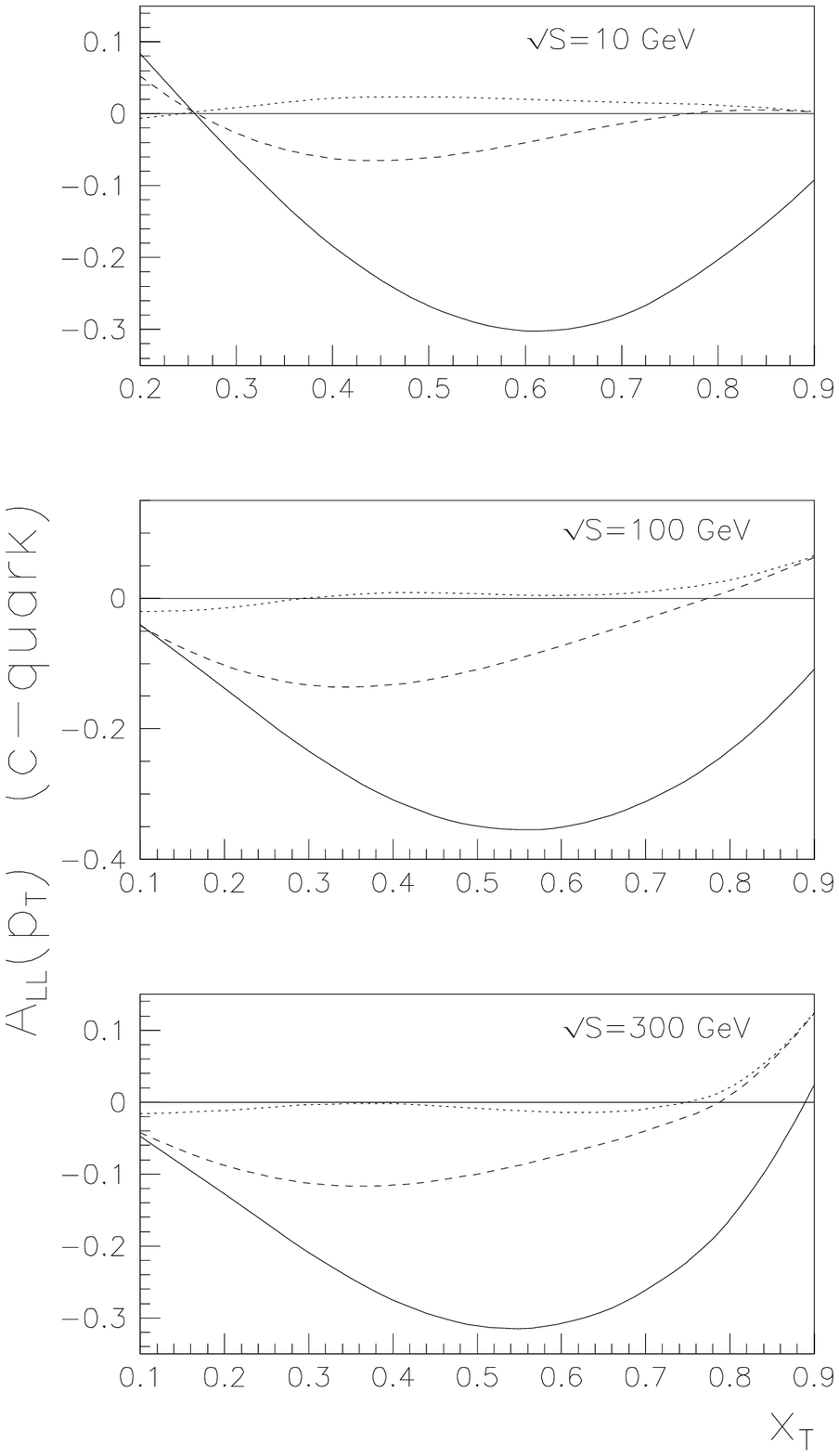}}
\mbox{\epsfysize=86mm\epsffile{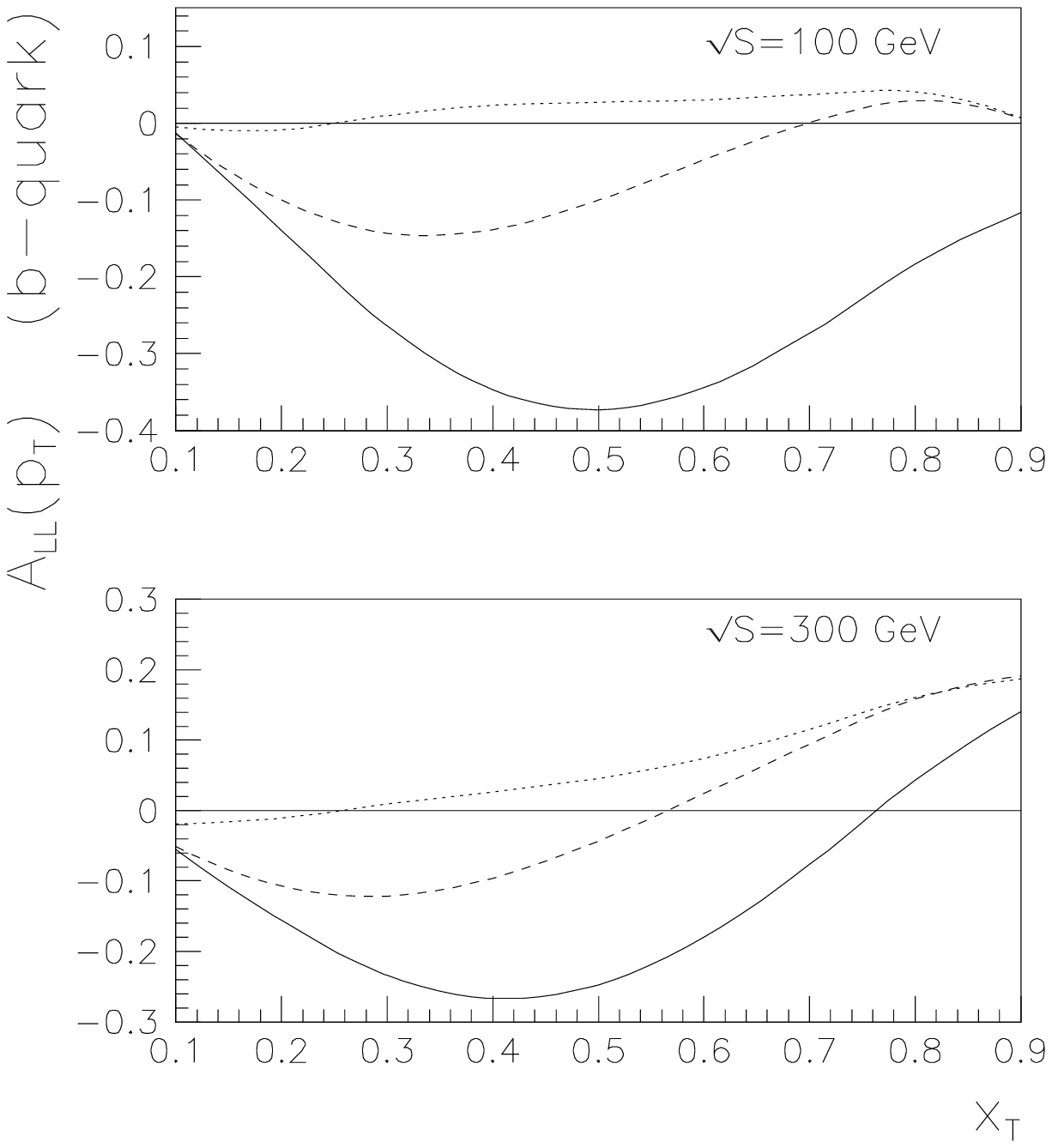}}
\caption{\small Differential asymmetry as defined in the text for three
sets of
GS polarized pd. Solid lines: set A, dashed: set B, dotted: set C.}
\label{F1}
\end{figure}
For the unpolarized cross sections we use the pd of \cite{CTEQ}. Our figures
are for three energies: $\sqrt{S}=10$ GeV (COMPASS and SLAC) 
and $\sqrt{S}=100$ and $300$ GeV (HERA).

Fig. 1 presents our NLO $A_{LL} \left( p_T \right)$ for $Q=c$ and $Q=b$
quark. Clearly, in general, sets A, B and C lead to $\Delta d\sigma/dp_T$' s
well separated. Also note that, in general, at small  $x_T$' s
$A_{LL} \left( p_T \right)$ are relatively small; exception is $\sqrt{S}=10$ 
GeV, when $A_{LL} \left( p_T \right)$ stay relatively sizable. On the
other hand, at low $x_T$ both the polarized and unpolarized cross sections
are relatively large; this is particularly true for the unpolarized ones
at HERA. Then, at HERA, at small $x_T$, $A_{LL} \left( p_T \right)$ are
very small, as was indeed found in \cite{BS}.

\begin{figure}
\centering
\mbox{\epsfysize=86mm\epsffile{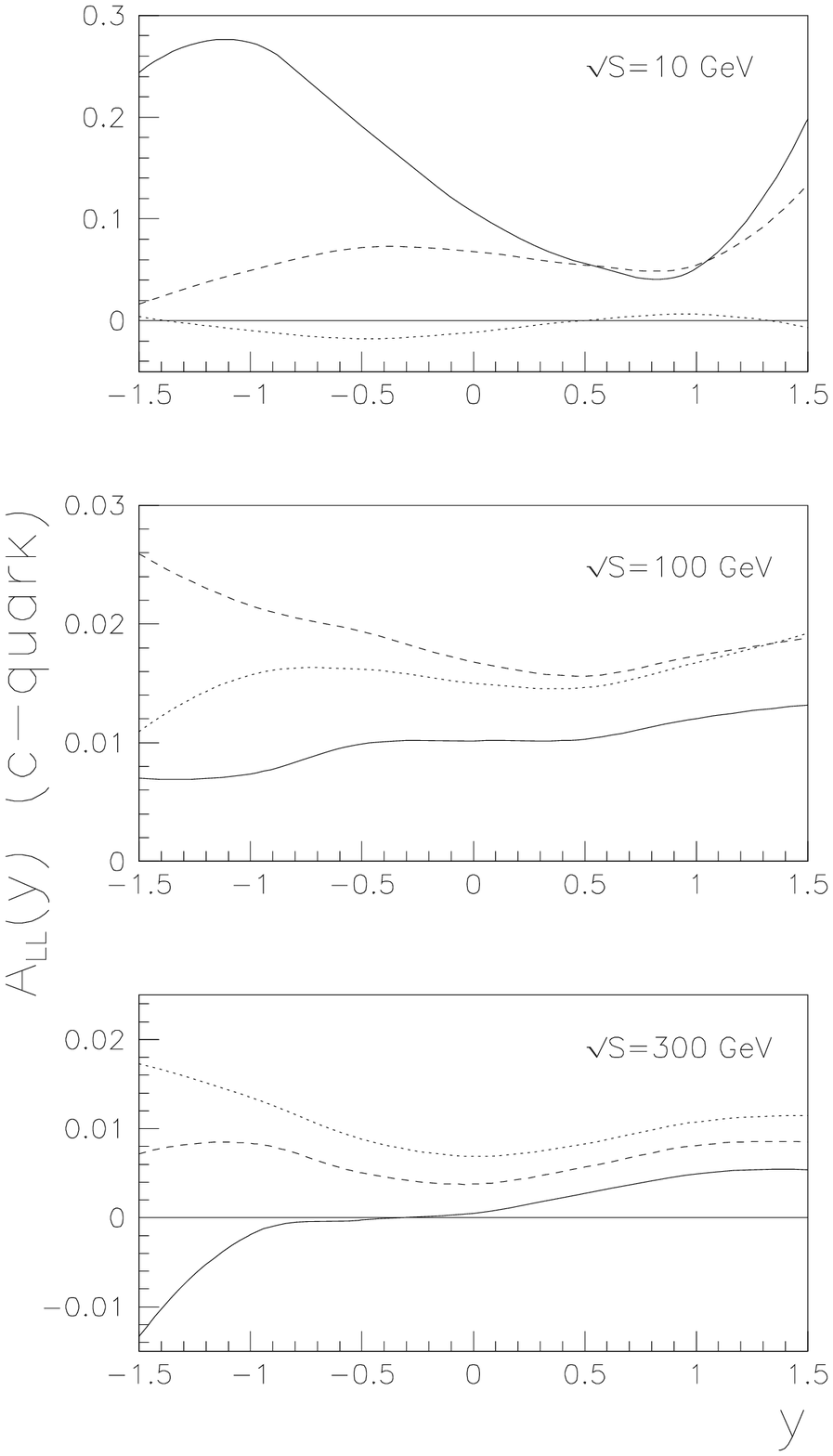}}
\mbox{\epsfysize=86mm\epsffile{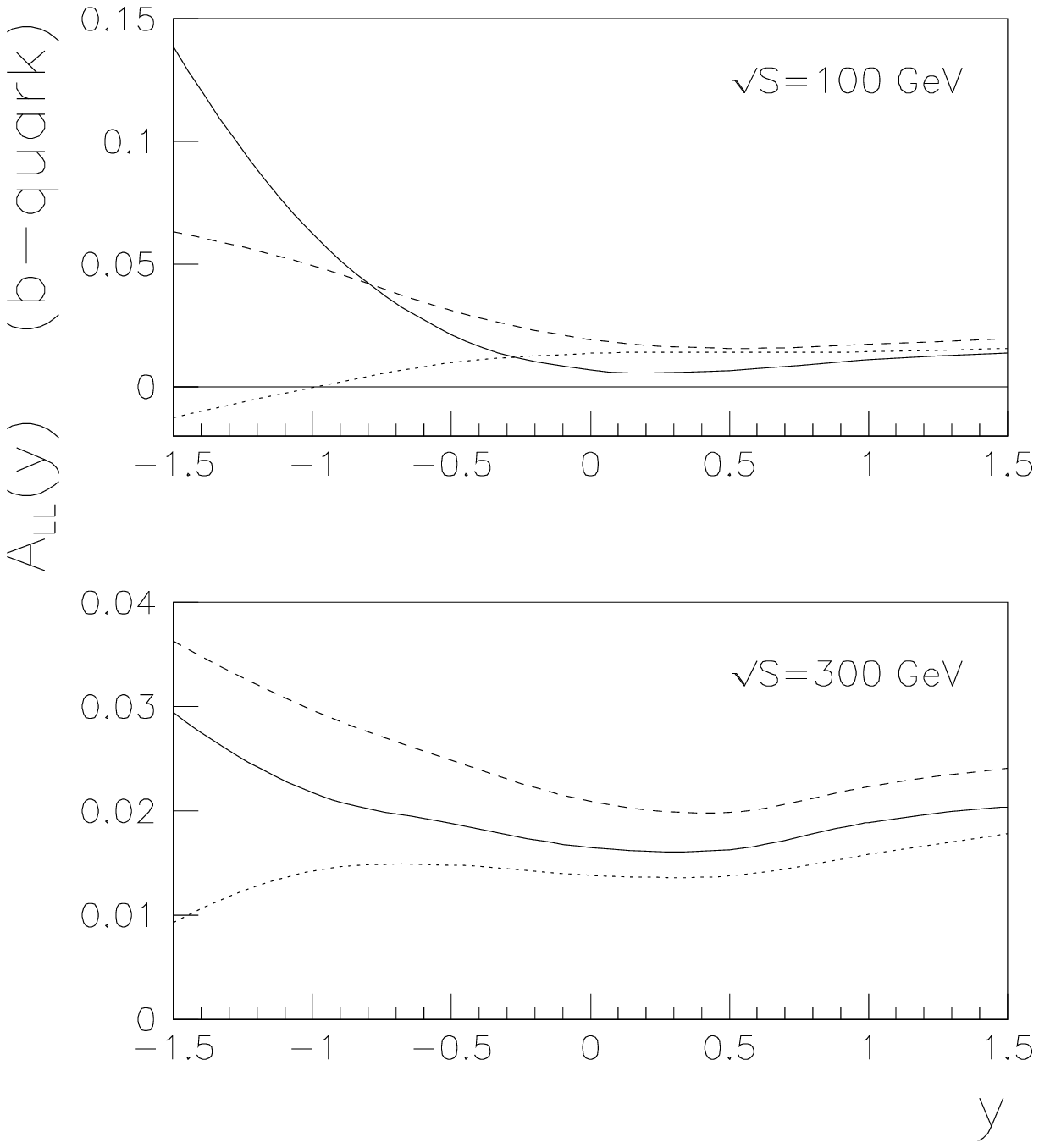}}
\caption{\small Rapidity differential asymmetry as defined in the text for
three
sets of GS polarized pd. Lines as in Fig.~1.}
\label{F2}
\end{figure}  

Fig. 2 presents $A_{LL} \left( y \right)$ and Fig. 3 the sizes of 
$\Delta d\sigma/dp_T$ and $\Delta d\sigma/dy$. In the later we do not plot
results for $\sqrt{S}=300$ GeV as the cross sections are too small to be
measured. For c-quark it seems that $\Delta d\sigma/dp_T$
is measurable up to $x_T \simeq 0.4$. Also, at $\sqrt{S}=10$ GeV, 
looking at Fig. 2 and Fig. 3 we see that in the range $1 \leq y \leq 1.5$
\textit{both} $A_{LL} \left( y \right)$ and $\Delta d\sigma/dy$ are quite
sizeable, so useful information on $\Delta g$ could well be obtained.

$\Delta d\sigma/dy$ of Fig. 3 show also that heavy quarks are mainly produced
forwards with respect to the (incoming) photon.
 
As we discuss in detail in \cite{MCG}, in general, we agree with \cite{BS}.
Nevertheless, perhaps on the basis of the smallness of HERA $A_{LL}$
at small $x_T$, Ref. \cite{BS} concludes that HERA is rather useless in
specifying $\Delta g$. Here we argue that it may  be not so: On the basis 
of Fig. 1, reconstruct events and select only those with, say, $x_T > 0.2$; 
in other words, carry integrations of $\Delta d\sigma/dp_T$ over some cut
phase space.
This may well enhance the resulting $A_{LL}$.

Of course, to reach a definitive conclusion, more work is needed, including 
an estimate of the corresponding errors.

\begin{figure}
\centering   
\mbox{\epsfysize=86mm\epsffile{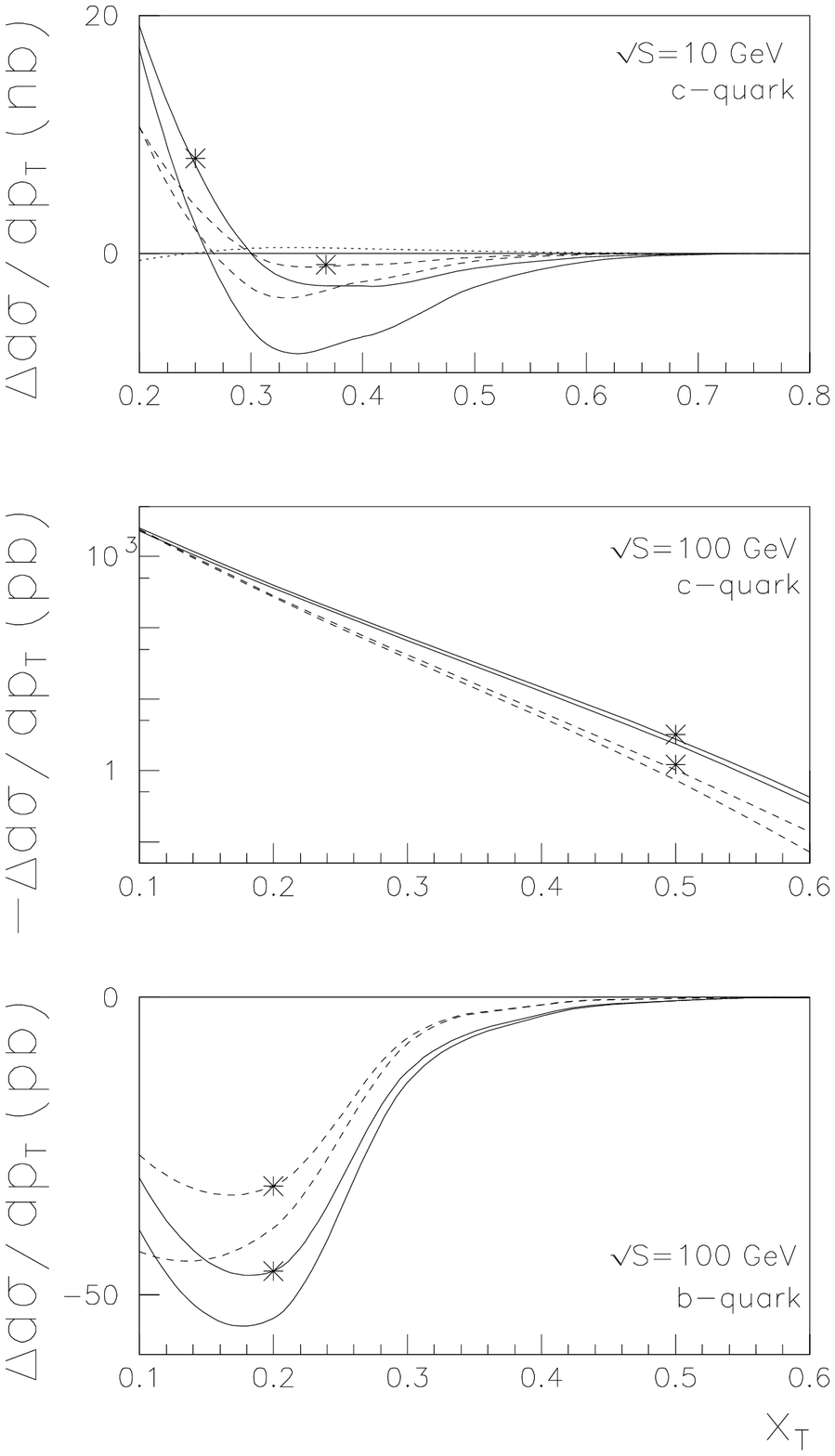}}
\mbox{\epsfysize=86mm\epsffile{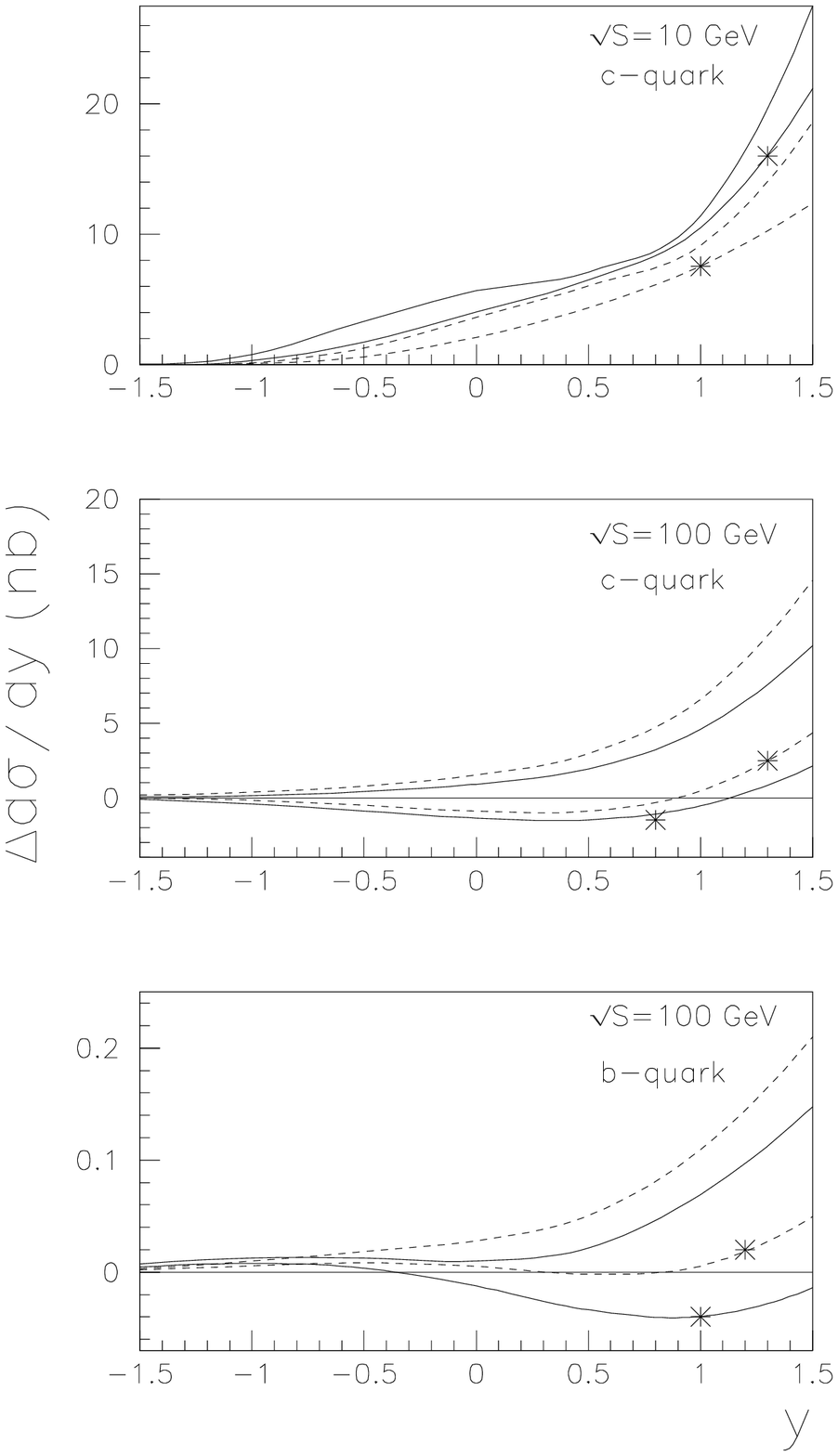}}
\caption{\small Polarized cross sections for two sets of GS polarized pd.
Solid lines: set A, dashed: set B, dotted: contribution due to the light
quark subprocess (\protect\ref{gamq}). Leading order results are marked
with stars.}
\label{F3}
\end{figure}

\bigskip
{\small
We would like to thank V. Spanos and G. Veropoulos for participating
in part of the calculation.
Z.M. would like to thank the Organizing Committee of the
"Praha-Spin99" Workshop, personally A. Efremov and M. Finger, for inviting
him to give a talk in this wonderful city. As well, Z.M. thanks participants
of the Workshop for interesting discussions. This work has been supported
by the Natural Sciences and Engineering research Council of Canada and
by the Secretariat of Research and Technology of Greece}
\bigskip


\end {document}